# The Hardy Experiment in the Transactional Interpretation

Ruth E. Kastner[†]

version of 6 August 2010

ABSTRACT. The Hardy experiment is analyzed from the standpoint of the Transactional Interpretation (TI) in its possibilist variant, PTI. It is argued that PTI provides a natural and illuminating account of the associated phenomena, resolving the apparent paradox which arises from the mistaken notion that components of the state vector labeled by spacetime regions imply the actual presence of a corpuscle in that region.

1. Introduction.

The Hardy experiment (Hardy 1992a) presents an apparent paradox based on a combination of two "interaction free measurements" of the kind proposed by Elitzur and Vaidman (1993). In such measurements, a Mach-Zehnder interferometer (MZI) is tuned so that one of the two detectors, a "silent detector" typically labeled D, will never be activated unless there is an obstruction in one of the arms (referred to in what follows as the 'blocking' arm and labeled *w,* see Figure 1). The Hardy experiment uses two overlapping MZIs, one for an electron and the other for a positron. Hardy's idealized presentation assumes that if the electron and positron meet in the overlapping region corresponding to both blocking arms, they will annihilate with certainty. (This is not strictly speaking correct, of course, since there is an amplitude less than unity for electron-position annihilation into two photons to occur no matter how close the particles get[1].) It turns out that even in cases where there is no annihilation (i.e., both particles are 'not in' the overlapping arms corresponding to the term |w+, w–>), both detectors D+ and D- can activate. This outcome theoretically occurs with a probability of 1/16. If we think of quanta as having definite whereabouts in the apparatus at all times, this seems paradoxical, since each individual apparatus is supposedly only able to have its D detector activated when something is blocking arm *w*. The event of both detectors D activating therefore seems to imply that both quanta must be in arms *w+* and *w-*, but then they should annihilate (or at least be mutually scattered), so presumably could not reach the detector area at all. Hence the paradox.

However, the above is only a paradox if we insist on thinking of quantum objects as classical corpuscles carrying energy and momentum along specific trajectories. This classical "billiard ball" story mistakenly tells us that an amplitude for an interaction to occur somewhere (e.g., in the blocking arm of the MZI) means that a corpuscle must actually be physically present there if some other detection (e.g., at D) occurs which depends on the given amplitude. This "billiard ball" notion is what is denied in the transactional interpretation (TI) (cf. Cramer (1986) and (2005)). In TI, quanta are not corpuscles pursuing trajectories. Amplitudes describe offer and confirmation waves which themselves do not transfer energy, but which can give rise to transactions. It is the completed (actualized) transactions that transfer energy and other conserved quantities, and which can therefore activate detectors.

---

[†]University of Maryland, College Park; Foundations of Physics Group. rkastner@umd.edu
[1] Cf. Berestetskii, Lifshitz and Pitaevski (2004), pp. 368-70.

To review the basics of TI: offer and confirmation waves ("OW": $\Psi(x)$ and "CW": $\Psi^*(x)$) are solutions of the Schrödinger equation and its complex conjugate, respectively. The emitter emits an OW and one or more absorbers (indexed by $i$) generate CWs in response to the OW component received by them. (For details of this process, see Cramer 1986.) Under the possibilist version of TI presented recently, "PTI," (Kastner 2010), these are *real dynamical possibility waves* of the kind suggested informally by Heisenberg.[2] It needs to be emphasized, to avoid misunderstanding, that such possibilities are not taken to be all-encompassing 'possible worlds' in a Lewisian sense[3] nor modal analogs of Everettian worlds which diverge into separate dynamical realms, but simply possibilities for *particular* transfers of conserved quantities in a *single* actualized spacetime.

Transactions are observer-free collapses described by the Born Rule. They occur stochastically based on responses of CWs to the initial OW from an emitter. The final amplitudes of the various CWs are described by the product $\psi_i^* \psi_i$. A CW response to an emitted OW is a necessary but not sufficient condition for a particular transaction to be actualized; the statistical weight of each possible, or incipient, transaction embodies the objective uncertainty of quantum outcomes. The weighted set of possible transactions for a particular experiment is what is described by the von Neumann "process 1" mixed state.[4]

TI thus treats absorption on the same dynamical footing as emission, providing an unambiguous account of how a 'measurement' is finalized, without the infinite regress of apparatuses or observers infecting the standard accounts of quantum measurement which neglect absorption. It is also harmonious with relativity (Cramer 1986, 668-9) and finds support for its even-handed treatment of emission and absorption in quantum field theory, which treats absorption and emission symmetrically.[5] (Emission can be said to be privileged only insofar as it is the starting point for a transaction.) Transactions are irreducibly stochastic collapses triggered by absorption events. So in TI, measurements--and any other empirically observable events--are just the results of transactions. There is no need to assign wave functions to macroscopic pointer coordinates, observers, or observer minds (nor, under TI, would this be correct--since an offer wave describes an unabsorbed possibility while macrosopic objects such as pointers and observers are conglomerates of actualized events based on completed transactions).

---

[2] E.g., "Atoms and the elementary particles themselves are not real; they form a world of potentialities or possibilities rather than things of the facts." (Heisenberg 1958/2007).
[3] As explicated in David Lewis' classic (1986).
[4] See, e.g., Bub (1997, 34).
[5] In this regard, note that the expression for a quantum field operator associated with a particular spacetime point is a sum of creation (emission) and annihilation (absorption) operators. Cf. Mandl and Shaw, p. 44. Emission of a particle is physically equivalent to absorption of an antiparticle, and *mutatis mutandis*. Absorption is just as important as emission in relativistic theories. It is only in traditional nonrelativistic quantum mechanics interpretations that absorption is ignored; TI remedies that discrepancy. This is not to say that TI finds its best relativistic expression in terms of QFT; a more suitable approach is suggested by 'action at a distance' theories such as that of Davies (1970). In terms of PTI, "action at a distance" simply means that interactions do not occur *in spacetime* but in the realm of possibilities transcending spacetime. This could be seen as analogous to the causal connection of two distant spacetime points 'locally' by way

In what follows, the Hardy experiment will be described in more detail and then an account of the phenomena will be given from the standpoint of TI.

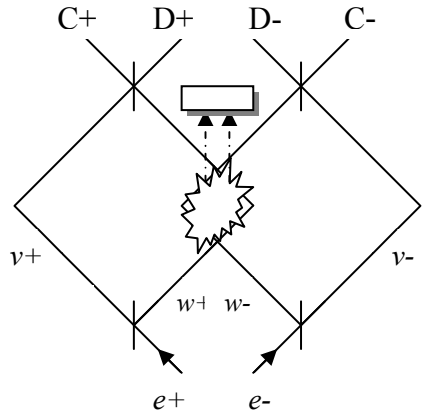

Figure 1. A schematic diagram of the Hardy experiment.

2. Details of the Hardy Experiment

The state of a quantum after passing the first beam splitter (a half-silvered mirror indicated in Figure 1 by a short vertical line) is[6]

---

of a wormhole. What seems like nonlocal connection from the standpoint of spacetime manifold becomes 'local' from the standpoint of a higher dimensional connection.

[6] Reflections result in a phase change of $\pi/2$, or a factor of $i$, for the component reflected.

$$|\psi_1\rangle = \frac{1}{\sqrt{2}}\big[|v\rangle + i|w\rangle\big] \tag{1}$$

Subsequently, each of the components $v$ and $w$ evolves as follows through the second beam splitter (the labels $c$ and $d$ refer to paths leading to the respective detectors C and D):

$$|v\rangle \to \frac{1}{\sqrt{2}}\big[|d\rangle + i|c\rangle\big] \tag{2a}$$

$$|w\rangle \to \frac{1}{\sqrt{2}}\big[|c\rangle + i|d\rangle\big] \tag{2b}$$

So that the state $|\psi_1\rangle$ evolves to $i|c\rangle$ when there is nothing obstructing either arm of the MZI.

The total system's state just after the first beam splitter is:

$$\begin{aligned}|\Psi_1\rangle &= \frac{1}{2}\big[|v+\rangle + i|w+\rangle\big] \otimes \big[|v-\rangle + i|w-\rangle\big] \\ &= \frac{1}{2}\big[|v+,v-\rangle + i|v+,w-\rangle + i|v-,w+\rangle - |w+,w-\rangle\big],\end{aligned} \tag{3}$$

where the kets with two labels are elements of the 4-dimension Hilbert space of the combined system.

The fourth term in (3) represents electron-positron annihilation in Hardy's idealization, which assumes that e+ and e− annihilate with certainty into two photons when they are both in the overlapping region. The two photons, indicated by the upward dotted arrows, are absorbed by a detector (indicated by the shadowed rectangle). If the two quanta were non-annihilating objects, such as two (coherent[7]) photons, the total state would simply evolve to $-|c+,c-\rangle$ and all quanta would be detected at C+,-. However, with the fourth term absent (i.e., according to the idealization, in cases where the electron and positron do not annihilate), we need to follow the evolution of the remaining three terms to see what detections are possible. Considering only the amplitudes of the component $|d+,d-\rangle$ for times after the second beam-splitter, we find the following contributions from the first three terms in (3):

---

[7] The two quanta have to be perfectly in phase for cancellation to occur.

$$|v+,v-\rangle \quad \text{gives} \quad \frac{1}{4}|d+,d-\rangle \tag{4a}$$

$$|v+,w-\rangle \quad \text{gives} \quad -\frac{1}{4}|d+,d-\rangle \tag{4b}$$

$$|v-,w+\rangle \quad \text{gives} \quad -\frac{1}{4}|d+,d-\rangle \tag{4c}$$

(and note that, if annhilation were not possible, the fourth term would give the same contribution as $|v+,v-\rangle$, thus canceling all contributions of $|d+,d-\rangle$).

Thus the fact that the three remaining terms contribute a nonzero amplitude for $|d+,d-\rangle$ (specifically, an amplitude of ¼) makes the detection at D+,D- possible when discounting contributions from the term $|w+,w-\rangle$, which leads to annihilation. Now let us see how TI describes this experiment.

3. The TI account

First, recall (e.g., Cramer 1986) that according to TI, transfers of energy resulting in detection occur only as a result of actualized transactions (as reviewed in Section 1). Yet there is much that goes on 'before' a transaction can occur.[8] The following are necessary (but not sufficient) conditions: first, an offer wave (OW) is emitted. In the Hardy experiment there are two single-quantum OWs corresponding to the electron and positron. The OWs propagate until they encounter a possible absorber. Thus the first opportunity for absorption corresponds to the term $|w+,w-\rangle$, in which the two OWs may encounter each other. As mentioned earlier, an accurate treatment of this situation would consider the relativistic scattering cross-section for e+, e- annihilation, but let's restrict the discussion to the nonrelativistic idealization presented by Hardy and assume that a realized transaction corresponding to this term is equivalent to annihilation. In this case, an *incipient* (possible) transaction is established in virtue of confirmation waves generated by the detector for the photons.[9] The generation of confirmation waves is a necessary condition for a transaction, but as noted above, not sufficient. The 'choice' of which transaction is realized is irreducibly stochastic (i.e., there is no determinate sufficient condition for a transaction). The probability of the annihilation transaction is given by the product of the OW and CW amplitudes, or ¼ (see Cramer 2005 for a quick introduction).

---

[8] In this context, 'before' means a-spatiotermporally, in the proposed possibility space, not 'at an earlier time index' in spacetime. This is roughly analogous to Cramer's 'pseudotime,' except that that was a heuristic term; whereas the possibility space discussed here is considered to be a genuine physical substratum beyond spacetime. In this regard, it should be noted that Stapp (1979) has previously proposed that the order in which events come into being need not be identified with their spatiotemporal ordering. This is applicable to the current proposal, in which events are actualized via transactions which place them into ordered relations with other transacted (actualized) events in spacetime. "Before" thus refers to an ontological order, not a spatiotemrporal order.

[9] For details on how this is treated relativistically (i.e., in the context of particle creation and destruction), see Chiatti (1994). He proposes a time-symmetric Feynman sum-over-paths approach encompassing particle interactions such as e+e- anniilation with clearly defined "in" and "out" states. The confirmation is assumed to be generated based on the final 'out' states; in this case, the two photons.

If this annihilation transaction does not occur, there is still an OW component for the combined system corresponding to a $|d+, d-\rangle$ transaction, with an amplitude of ¼. (see (4)). That is, OWs for each quantum (e+ and e-) can reach detectors D+,-. The detectors are composed of absorbers which respond to each OW by returning a CW of the same amplitude to the respective emitters (e+ and e-). The probability of this transaction is the final amplitude of this CW, which undergoes the same attenuation (through interactions with components of the apparatus such as beam splitters; see Cramer 1986, pp 661-2, 674-5) as the original OW; the final amplitude is given by the Born Rule, (¼) (¼) = 1/16.

Thus there is nothing paradoxical if we see these processes as involving interactions between offer and confirmation waves rather than as dictating the supposed whereabouts of localized particles. There is, however, one challenge for the TI account, similar to the challenge presented for TI in the Quantum Liar Experiment of Elitzur and Dolev (2006), and addressed in Kastner (2010) . This challenge consists in the following. The standard TI account talks about OW and CW as propagating in spacetime through experimental apparatuses, but experiments like the one discussed in this paper and in the "Quantum Liar" show that this is an oversimplification. These experiments involve cases in which a transaction corresponding to a particular term may not occur, but an OW component appearing in that 'untransacted' term is still needed for other possible transactions that could still occur. In the Hardy experiment, this situation occurs because the single-quantum components corresponding to the "blocking" arm, $|w+\rangle$ and $|w-\rangle$, are still needed for transactions involving detectors C and D. The latter possibilities arise from the terms (4b) and (4c).

So we can't just say that if the annihilation transaction corresponding to the term $|w+, w-\rangle$ doesn't occur, then the entire content of that term is 'out of the picture', because we still need its single-particle components. If we try to visualize single-particle waves propagating through the apparatus, we end up with an awkward account in which (for example) the positron OW component $|w+\rangle$ 'decides' not to engage in a transaction placing it in the blocking arm (corresponding to $|w+, w-\rangle$), but still has to be present in the blocking arm (corresponding to $|v-, w+\rangle$) to end up with a component that can be absorbed by D-. Does the positron OW "go back home and try again" after the failed $|w+, w-\rangle$ transaction?

The resolution of this puzzle is the same as presented in Kastner (2010), and involves a paradigm change in the relationship of quantum objects to spacetime. Specifically, we cannot picture the entities described by quantum states as literally propagating *in* spacetime through the arms of an MZI. Instead, it is proposed that quantum states describe dynamical *possibilities* whose domain is mathematically described by Hilbert Space, not 3-space or spacetime. Spacetime is the theater of completed transactions, not the domain of quantum states which requires a larger mathematical structure (because there are enormously more possibilities than can be actualized in spacetime[10]). An accurate description of an experiment involving microscopic systems which require a quantum mechanical description must treat the *entire* experimental apparatus and quantum system as a nexus of OW, CW, incipient transactions, and actualized transactions. The macroscopic features of the apparatus will correspond to highly probable and persistent transactions which enable it to be thought of as 'existing in spacetime,' since spacetime is the domain of the structured set of actualized (successful) transactions. However, elements of the experiment with significantly fewer and less probable transactions (the electron and positron in this case) do not really exist in spacetime but interact with the relevant aspects of the apparatus (i.e., absorbers) on the level of possibility (OW and CW), in *the larger possibility space* corresponding to all the quanta comprising the entire system. At this point, it is worthwhile to recall Heisenberg's famous comment quoted here in footnote 2.[11]

The foregoing implies, as noted in Kastner (2010), that such possibilities are not just conceptual abstractions but entities that exist in a pre-spacetime realm. (For a similar proposal concerning the necessity for a 'pre-spacetime' realm, see Elitzur and Dolev 2005). The modal realist version of TI thus proposed is "possibilist TI", or PTI. The overall picture is one of macroscopic objects, such as experimental devices and their detection phenomena, being 'bootstrapped up" from transactions rooted in a larger space of possibilities. We cannot directly observe that larger space, but can infer it from the fact that the quantum entities that give rise to spacetime phenomena must be described by a larger mathematical space. Again, as observed in Section 1, this is *not* just a modal realist version of the Everett Interpretation. The possibilities in PTI are for specific events, and there is no 'branching' in which various actualized versions of events goes their own separate ways. There is only one actualized reality here: the spatiotemporal realm emergent from transactions.

If this picture strikes one as farfetched and/or metaphysically extravagant, it must be pointed out that competing interpretations such as Bohm's theory and Everettian or Many Worlds Interpretations (MWI) could, by the same standard, be similarly be considered farfetched and/or metaphysically extravagant. While a detailed review of these interpretations is beyond the scope of this paper, it must at least be noted that a Bohmian interpretation of interaction-free measurements (IFM) such as the Hardy experiment involves the notion of 'empty waves' (i.e., wave packets without Bohmian particles; cf. Hardy 1992b). Multiparticle Bohmian guiding waves are subject to the same concern about the relationship of multiparticle quantum states to spacetime as are TI's OW and CW, since the Bohmian waves are elements of a multidimensional configuration space, not waves in spacetime. If one wishes to be skeptical about the relationship of multiparticle offer and confirmation waves to experimental devices, for consistency one needs to apply the same skepticism to multiparticle Bohmian 'empty waves' which are alleged to propagate through the apparatus but which do not 'fit into' ordinary spacetime any more than the offer and confirmation waves of TI.

---

[10] An interesting image reflecting this mathematical fact can be found on the cover of Bub's *Interpreting the Quantum World* (1997). The cover image shows M. C. Escher's famous print "Waterfall," depicting a scene with a physically impossible topology (i.e., one that could not actually fit into spacetime). Three separate areas of the print are highlighted, and each of these could exist in isolation in spacetime, but the global connections between them cannot. In the interpretation proposed here, the smaller highlighted 'normal' areas represent the actualized transactions, and the larger shaded area of topologically 'impossible' global interconnections belong to the pre-spacetime realm of possibility (i.e., offer and confirmation waves).

[11] By "not real," Heisenberg meant that they do not exist in spacetime. Being a physicist, he assumed that real things must exist in spacetime. The approach here is to suggest that the possibilities represented by quantum states can be considered real in virtue of their physical efficacy as described herein.

As for MWI, again, a detailed critical review of Everettian approaches is beyond the scope of this paper, but a good place to start is Kent (2010). If the putative splitting of worlds occurs in a larger space called a "multiverse" (Deutsch, 1998), then this provides equal ontological license for offer and confirmation waves to propagate in an analogous larger space. TI carries the additional benefit of a straightforward physical explanation of the Born Rule clearly unavailable in MWI, as argued in Kent (2010).

4. Conclusion

The Hardy experiment has been analyzed from the standpoint of the Transactional Interpretation (TI) in its possibilist variant, PTI. It has been argued that PTI provides a natural and illuminating account of the associated phenomena, resolving the apparent paradox which arises from the mistaken notion that components of the state vector labeled by spacetime regions imply the actual presence of a corpuscle in that region. Instead, such components refer to possibilities that may or may not be actualized in a transaction. Elitzur and Dolev (2001) reach a similar conclusion (i.e., that the notion of a spacetime trajectory for quantum objects must be abandoned) in the context of a variation of Hardy's experiment (Hardy 1992b). They describe seemingly bizarre behavior by the wave function when encountering a sequence of possible absorbers and conclude: "Ordinary concepts of motion, which sometimes remain implicit within prevailing interpretations, are in adequate to explain this behavior...The most prudent description of this result is that a wave function, when interacting with a row of other wave functions one after another, does not seem to comply with ordinary notion of causality, space and time." (2001, 6).

The metaphysical picture proposed here is that the experimental apparatus seems persistent in virtue of the highly probable and frequent transactions comprising it, while the quantum phenomena seem less so because their transactions are far fewer and much less probable. Yet the entire phenomenon (apparatus + quantum system) is rooted beyond spacetime, in the domain of possibilities described by Hilbert space, and that is where the fundamental interactions (involving exchanges of offer and confirmation waves) take place. It has been observed that Bohmian 'empty wave' interpretations of IFMs such as the Hardy experiment are subject to exactly the same challenge concerning the relationship of multiparticle quantum states or wave functions to ordinary spacetime, and that this puzzle is addressed squarely by PTI while it apparently has yet to be recognized in Bohmian accounts of the Hardy experiment. In addition, the 'larger space' required for a consistent account of TI's offer and confirmation waves has precedent in the notion of 'multiverse' in many-worlds interpretations and therefore, in order to avoid a double standard, should not be rejected as metaphysically extravagant.

Acknowledgements.

The author is grateful for numerous valuable suggestions by an anonymous referee.